# Universal and generalizable restoration strategies for degraded ecological networks

**Udit Bhatia**

Sustainability and Data Sciences Lab, Department of Civil & Environmental Engineering, Northeastern University, Boston, MA 02115, USA.

E-mail: bhatia.u@husky.neu.edu

**Tarik C. Gouhier**

Department of Marine and Environmental Sciences, Marine Science Center, Northeastern University, Nahant, Massachusetts 01908, USA.

E-mail: t.gouhier@neu.edu

**Auroop R. Ganguly***

Sustainability and Data Sciences Lab, Department of Civil & Environmental Engineering, Northeastern University, Boston, MA 02115, USA.

E-mail: a.ganguly@neu.edu

* Corresponding Author Contact: Sustainability and Data Sciences Lab, Department of Civil & Environmental Engineering, 400 Snell Engineering, Northeastern University, 360 Huntington Avenue, Boston, MA 02115, USA; Phone: +1-617-373-6005; Email: a.ganguly@neu.edu; Internet: http://www.civ.neu.edu/people/ganguly-auroop

**Humans are increasingly stressing ecosystems via habitat destruction, climate change and global population movements leading to the widespread loss of biodiversity and the disruption of key ecological services[1–3]. Ecosystems characterized primarily by mutualistic relationships between species such as plant-pollinator interactions may be particularly vulnerable to such perturbations because the loss of biodiversity can cause extinction cascades that can compromise the entire network[4,5]. Here, we develop a general restoration strategy based on network-science for degraded ecosystems. Specifically, we show that network topology can be used to identify the optimal sequence of species reintroductions needed to maximize biodiversity gains following partial and full ecosystem collapse[6,7]. This restoration strategy generalizes across topologically-disparate and geographically-distributed ecosystems. Additionally, we find that although higher connectance and diversity promote persistence in pristine ecosystems[8], these attributes reduce the effectiveness of restoration efforts in degraded networks. Hence, focusing on restoring the factors that promote persistence in pristine ecosystems may yield suboptimal recovery strategies for degraded ecosystems. Overall, our results have important insights for designing effective ecosystem restoration strategies to preserve biodiversity and ensure the delivery of critical natural services that fuel economic development, food security and human health around the globe[9,10].**

Environmental change has led o the widespread loss of species around the world[1,4,11]. This type of ecosystem degradation is particularly dangerous for mutualistic networks because their stability is dependent upon the maintenance of species diversity and key interactions[8,12]. Hence, the loss of a species can ripple through the mutualistic network and cause a cascade of secondary extinctions that could further compromise the stability of the entire ecosystem [6,13]. These properties suggest that the recovery of ecosystems may be optimized by sequentially restoring the most critical species and their interactions in the network [14,15]. A key question is whether such a strategy can be generalized across disparate ecosystems. Specifically, can a universal criterion be used to identify the criticality of a species despite large differences in the complexity, topology and geographical distribution of mutualistic networks [16]?

A generic way of defining the criticality of species is to measure their 'keystoneness' or the degree to which their impact on ecosystems is large relative to their abundance[16,17]. Although initially developed in relatively small predator-prey communities[18–20], this operational definition of keystoneness has been extended to complex networks based on species' functional or topological importance[21,22]. However, there is a lack of consensus in (i) how to classify the extent to which a species can be considered a keystone based on its functional versus topological characteristics, (ii) whether the concept of keystoneness can be applied to disparate ecosystems, and (iii) the degree to which keystone species contribute to disproportionate biodiversity gains within complex network. This lack of consensus has hindered the development of universal and optimal restoration strategies for environmentally-degraded ecosystems.

Prior attempts to identify optimal restoration strategies have highlighted a variety of viable approaches. For instance, some studies have shown that closely interacting species, measured via closeness centrality, play a pivotal role in promoting network-wide robustness and persistence[22,23]. Closeness centrality has thus been suggested as a measure of species keystoneness based on its importance in understanding and ultimately preventing cascading secondary extinctions. Other approaches have focused on developing restoration strategies based on the response of the whole network to perturbations. Indeed, multiple previous studies have hypothesized that the optimal recovery strategy should be the mirror image of the sequence of species loss that generated the maximum number of secondary extinctions[14]. Additionally, the structure of mutualistic networks has been linked to their robustness to perturbation. For instance, higher connectance and diversity have been shown to promote persistence and resilience, defined as the speed at which a perturbed community returns to equilibrium[8]. However, despite these insights, a generic framework for identifying an effective overall recovery strategy remains elusive.

We conducted a systematic analysis of real plant-pollinator systems around the globe to identify a universal and optimal restoration strategy. Since identifying the optimal set of nodes that has the greatest impact on network recovery or disruption is a non-deterministic polynomial time hard problem[24], we identify the optimal restoration strategy based on network topology alone (note that our definition of "optimal" is not based on permutations of all possible node sequences, which is known to be practically infeasible and computationally intractable[24,25]; see Methods). Specifically, we examined whether restoration strategies based on keystoneness as measured by network centrality systematically resulted in larger gains in biodiversity compared to random or non-strategic restoration interventions. To do so, we used a dynamical model to simulate recovery via species reintroductions following perturbations ranging from partial (25% species loss) to complete (100% species loss) ecosystem collapse. We enforced obligate mutualism in that species were considered extinct following the loss of all their mutualistic partners[4].

We simulated four restoration strategies: generalism(degree), eigenspecies, betweenness and closeness centralities. We focused on these approaches because they can easily be applied to any mutualistic network regardless of ecological and environmental context[26]. Furthermore, the concept of network centrality has been applied to investigate the topological importance of plants and pollinators in relation to their degree of generalism[23]. Hence, such approaches have the greatest potential for yielding system-agnostic or context-independent restoration strategies. To quantify how network-based recovery strategies performed compared to naïve approaches, we used a null model in which species were restored randomly until complete recovery was achieved. To understand the relationship between network architecture and recovery from perturbations, we measured the following attributes: asymmetry (ratio of pollinators to plants), connectance (proportion of realized species interactions) [27], nestedness (degree to which specialists interact with a subset of species that interact with generalists) [28], modularity (degree of compartmentalization) [29] and diversity (number of nodes in the network). Many of these architectural characteristics have been linked to resilience in both mutualistic and non-mutualistic networks [8,30].

Our analysis of 39 plant-pollinator networks reveals that these systems are geographically and topologically distinct. This diversity and complexity is evident from the large variation in architecture between the networks as shown in Figure 1. Hence, a restoration strategy that is effective for these ecologically- and environmentally- disparate networks is likely to be general. To test the effectiveness of each restoration strategy, we computed the recovery score by measuring the size of the largest connected cluster in the network following species reintroductions (Extended Data Fig. 1). The recovery scores obtained from each strategy were converted to proportional Marginal Recovery Scores (MRS) by subtracting them from the median recovery score of the null (random) model. Since magnitudes of marginal recovery scores can depend on which species are going extinct (and subsequently restored) and to what extent systems are perturbed, we generated ensembles of 1,000 different sequences of species loss for the three perturbation scenarios (75%, 50% and 25% species loss). We examined (i) Marginal recovery scores (MRS) and (ii) the fraction of species that needed to be restored (FSR) in order to achieve peak marginal recovery scores.

Figure 2 shows that all restoration strategies yield similar modal patterns across the 39 networks, with large peaks in marginal recovery scores occurring for relatively low fractions of species restored (< 0.2). Despite these overall similarities, there were clear differences in terms of effectiveness between the strategies (Figure 2; Extended Data Fig. 2-4). Indeed, restoration based on generalism and betweenness led to higher peak marginal recovery scores (Figure 3 a,b; Extended Data Fig. 5) compared to the two other strategies (Fig. 3 c,d). While these two centrality measures were not significantly different in terms of median restoration scores (Fig. 3, p-value > 0.05), significant differences were observed in terms of the fraction of species required to achieve the peak marginal recovery score (p-value <0.05; Inset Fig. 3). Strategies based on closeness and eigenspecies centrality also exhibited greater variability in the fraction of species restored required to attain peak marginal recovery compared to those based on generalism and betweenness centrality (Fig. 3).Our results thus demonstrate that restoration strategies based on closeness centrality are suboptimal, a finding that clashes with previous theory [23]. Indeed, previous work has suggested that when perturbations propagate along the path of closeness centrality, biodiversity loss is likely to be fastest [23]. It has thus been hypothesized that the most effective recovery strategy should be the mirror image of the fastest destructive pathway [14]. Taken together, these two predictions thus suggest that closeness centrality is likely to represent the optimal restoration strategy. However, our results indicate that restoration based on closeness centrality is sub-optimal. This is likely due to the fact that species with high closeness scores interact with a small sub-community, which means that their reintroduction will result in slower gains in interactions and hence biodiversity.

We find that mutualistic network asymmetry is strongly related to marginal recovery, with networks characterized by high asymmetry exhibiting high peak marginal recovery scores (Fig. 4a). We also found significant associations between key measures of network structure and marginal recovery scores across all ecosystems. Nestedness, which is inversely related to nestedness temperature, was positively related to marginal recovery score (Fig. 4b). This suggests that both total species diversity and the relative size of the bipartitions systematically promote restoration success across all ecosystems. While researchers have demonstrated that higher diversity and connectance promote the persistence and resilience of mutualistic networks[13], our analysis reveals a significant negative relationship between connectance and marginal recovery scores (Fig. 4d). This could be because the greater redundancy of species interactions in networks characterized by high connectance leads to slower gains in biodiversity during the restoration process. Indeed, in such highly connected systems, restoration merely reinforces existing species interactions instead of forging new ones, which thus leads to smaller gains in biodiversity. No significant relationship was found between modularity and recovery scores for full and 75% species loss (Fig. 4c). However, positive relationships between these variables were observed for perturbations leading to the loss of 25% - 50% of species (See Extended Data Fig. 1).

Overall, our results suggest that certain network-based approaches are systematically more efficient for restoring degraded ecosystems characterized by distinct ecological and environmental contexts. Previous studies have primarily focused on understanding ecosystem robustness [14] and identified restoration strategies based on the functional traits of species. Although trait-based approaches can provide powerful insights for designing effective restoration strategies, the effort needed to acquire the data that underlie these types of analyses is non-trivial, especially given the large geographical, phylogenetic, ecological and environmental differences that often exist between ecosystems. Here, we have shown an alternative approach: the existence of a general and optimal restoration strategy based only on the network properties of the ecosystem. This approach not only identifies the optimal overall restoration strategy but the relative importance of each species at different stages of the recovery process. This suggests that the contribution of each species to network resilience varies depending on the level of ecosystem degradation.

Many scientists argue that we are either entering or in the midst of the sixth great mass extinction[2]. Intense human pressure, both direct and indirect, is having profound effects on natural environments. For example, global environmental change has severely disrupted critical mutualistic ecological networks[1]. Indeed, the loss of pollinators via habitat fragmentation could significantly affect the maintenance of plant diversity, ecosystem stability[8] and crop production[10]. Developing effective restoration strategies for perturbed ecosystems and designing interventions to reduce the damage from human or natural perturbations are crucial to address these issues.

**Acknowledgments** This research was primarily funded by NSF Cyber SEES (#1442728) as well as by NSF BIG DATA (# 1447587) and NSF Expeditions in Computing (#1029711). Data was obtained from http://www.web-of-life.es/ developed by Jordi Bascompte's lab . The authors thank Evan Kodra, CEO of risq corporation for valuable feedback.

**Author Contributions** UB, TCG and ARG designed the study and interpreted the results. UB analyzed and visualized the data. UB, TCG and ARG wrote the paper.

**Author Information** Reprints and permissions information is available at www.nature.com/reprints. The authors declare no competing financial interests. Readers are welcome to comment on the online version of the paper. Correspondence and requests for materials should be addressed to A.R.G. (a.ganguly@neu.edu)

# Methods

**Description of the model.** We constructed a regrowth model for degraded mutualistic networks to track how the number of species interactions *E* changes over time based on the sequence of species reintroductions prescribed by each restoration strategy (Extended Data Fig. 1). The general model can be expressed as:

$$E(t) = E(t-1) + f_i(t)D(i) - \sum_{k=1}^{t-1} a_{ik}$$

Where *E(t)* and *E(t-1)* represent the number of interactions present in the network at times *t* and *t-1* respectively, $f_i(t)$ is a function used to determine which species is reintroduced at time *t* based on each restoration strategy (see next section for details), *D(i)* is the degree (or number of interactions) of the $i^{th}$ species restored at time *t*, and $a_{ik}$ represents the number of interactions of species *i* in the network before time *t*. Number of interactions *(E(t))* controls the size of the largest connected cluster (LCC) in the network. Since LCC has been used to measure robustness of the real-life networks in ecological as well as non-ecological contexts [31], we use LCC as a measure of recovery score.

**Determining the sequence of species reintroductions and computing the marginal recovery score.** To determine the order in which species should be reintroduced, we considered network centrality measures that have been used to quantify the importance of nodes in both ecological and non-ecological contexts. Specifically, we chose (a) Degree (measure of species generalism), (b) Betweenness (measure of topological importance of species), (c) Closeness (species proximity to other species) and (d) Eigenspecies (importance of species because of its interactions) centrality. For each restoration strategy (a-d), we used the corresponding centrality metric to determine the identity of the species that should be reintroduced at time $t$ via $f_i(t)$.

We simulated four perturbation scenarios ranging from partial (25%, 50%, 75% species loss) to complete network collapse (100% species loss). To determine the robustness of our restoration strategy to the identity of the species removed from the ecosystem, we generated an ensemble of 1000 degraded networks by randomly selecting the species that were removed under the partial perturbation scenarios (i.e., 25%, 50%, and 75% species loss). We then computed the recovery score for each of perturbation scenario by calculating the median recovery score across the ensemble of 1000 degraded networks at each restoration step.

We used the random restoration of species as a null model to evaluate the performance of strategic intervention based on centrality scores. To do so, we computed an ensemble of 1000 random restoration sequences for each perturbation scenario and computed the median recovery score. We then defined the marginal recovery score as the differences between the strategic recovery score and the recovery score of null model at each restoration step (Extended Data Fig. 2-4)

We not that determination of an optimal solution based on a complete search of all possible permutation of node sequences is an NP hard (computationally intractable) problem [24,25]. This implies that the search for the true optimal solution becomes practically infeasible hence motivating (e.g. in our case) comparative evaluation of approaches based on topological and/or ecological attributes.

**Computing network attributes:** We computed a number of network attributes to determine whether they could be used to predict the recovery score of each ecosystem following perturbations ranging from partial to full species loss.

*Asymmetry*: We defined asymmetry as the ratio of the number of pollinators and plants in each mutualistic network. Highly asymmetric networks have asymmetry scores of greater than 3 or less than 1. Networks whose asymmetry scores fall between 1 and 3 are defined as symmetric or moderately asymmetric. In the case of obligate mutualism, each plant species is required to interact with at least one pollinator species[32]. In highly asymmetric networks, the rarer node type (e.g., plant or pollinator species) will thus become a limiting resource. Hence, we predicted that network asymmetry would be related to the recovery score.

*Nestedness*: Nestedness, a measure of structural organization in an ecosystem, was computed for all mutualistic networks. Specifically, we used nestedness temperature, which measures the degree to which the order of species extinctions is random. As nestedness temperature increases, the order of species extinctions becomes more stochastic. Conversely, as nestedness temperature decreases, the order of species extinctions becomes more deterministic. We focused on nestedness because previous studies found a positive relationship between the degree of nestedness and the persistence of mutualistic networks. We used the binmatnest program written by Miguel Rodríguez-Gironés to compute the nestedness temperature of bipartite networks.

*Modularity*: Modularity measures the degree to which a group of species interacts more among themselves than with species from other groups. This tendency thus results in the formation of densely connected sub-networks or modules. We used the walktrap algorithm to identify such sub-networks. This method identifies these sub-networks via random walks with the assumption that short random walks tend to stay in the same module or sub-network. The modularity of a given network is measured with respect to vertex types to measure how good the division is:

$$R = \frac{1}{2m} \sum_{i=0}^{n} \sum_{j=1;j\geq i}^{n} \left(A_{ij} - D_i \times \frac{D_j}{2m}\right) \times \delta(ci, cj), i, j$$

Where $A_{ij}$ is the element of adjacency matrix in row $i$ and column $j$, $D$ is the node degree, $ci$ and $cj$ is the type of node $i$ and $j$, respectively decided by community membership identified by walktrap algorithm. The sum goes over all pair of vertices and $\delta(ci, cj) = 1$ if $ci = cj$ and 0 otherwise. We focused on modularity because previous studies found a positive relationship between the degree of modularity and network stability (26).

***Connectance***: Connectance measures the proportion of realized interactions in a network. To measure connectance in mutualistic networks, we computed the total number of interactions between species divided by the total number of interactions possible between all N nodes. Mathematically, connectance can be expressed as:

$$C = \left(\sum_{i=0}^{n} \sum_{j=1;j\geq i}^{n} A_{ij}\right) / N^2$$

We also compute the connectance for bipartite networks, which can be expressed as:

$$C_{bi} = \left(\sum_{i=0}^{n} \sum_{j=1;j\geq i}^{n} A_{ij}\right) / (n \times m)$$

where $A_{ij}$ is same as equation (1); n and m represents the number of plants and pollinators in the network respectively.

**Statistical Analyses.** Statistical analyses are performed using Python 2.7. The statistical significance of the differences between mean value of peak MRS for different strategies is determined using Wicoxin test with Bonferroni correction. The null hypothesis used for comparison is that the peak of the MRS means are equal.

We use ordinary least squares (OLS) to estimate the relationship between the marginal recovery scores (MRS) and the various network attributes defined above. A relationship was deemed statistically significant if the associated *P*-value was less than 0.05. We also measured the degree of linear dependence between pairs of variables using the Pearson product-moment correlation coefficient. Results of statistical analyses for 25%, 50% and 75% perturbation scenarios are summarized in Extended Data Fig. 5-7

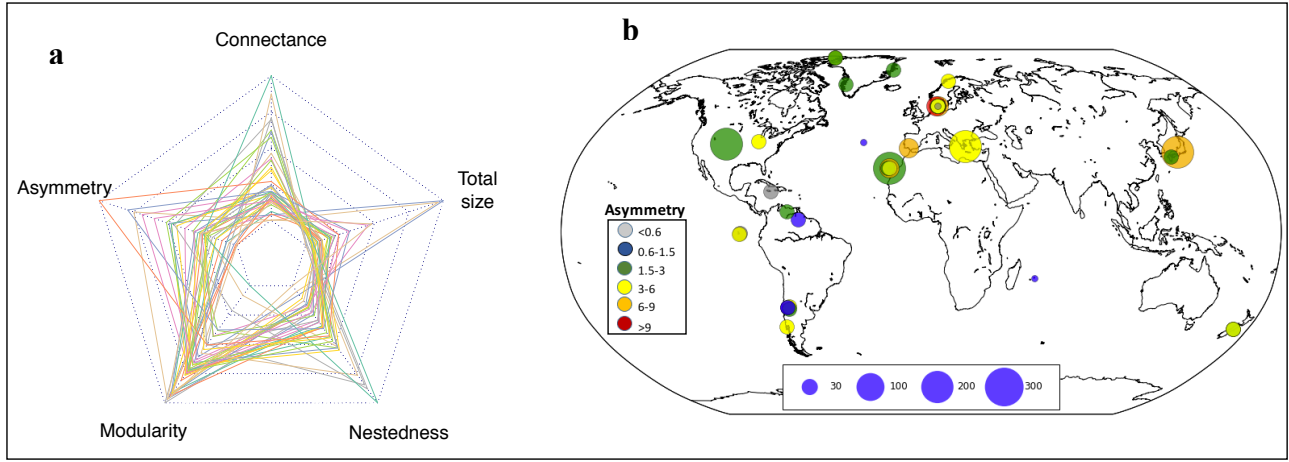

**Figure 1| (a)** Radar plot showing the network attributes for 39 plant-pollinator networks analyzed in our study. Length of spoke is proportional to the magnitude of the attribute relative to its maximum magnitude. **(b)** Geographical locations of the networks considered are shown. Node size indicates diversity within each network. Nodes colors represents the network asymmetry (ratio of pollinators to plants).

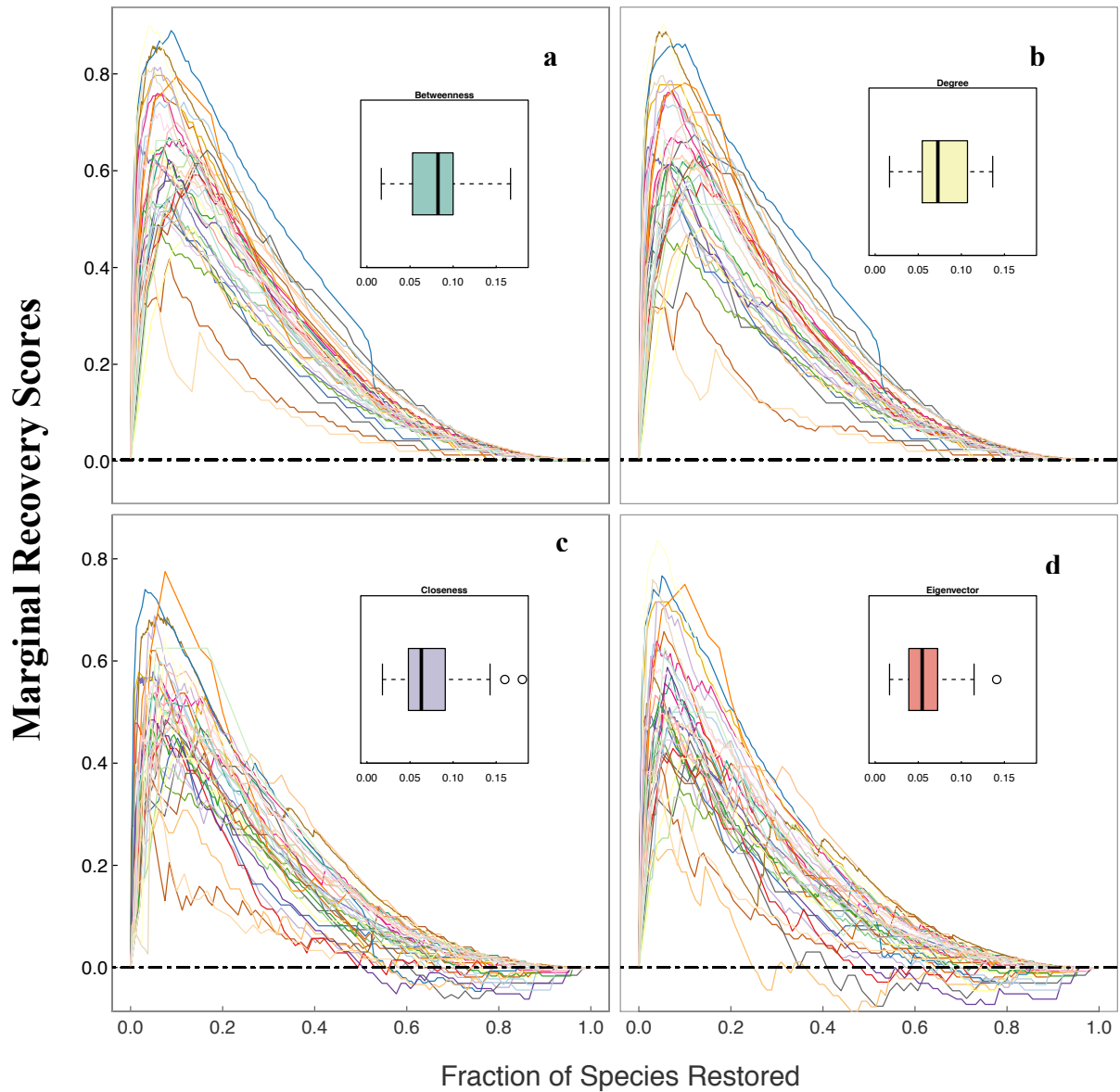

**Figure 2|** Marginal Recovery Scores (MRS) of full restoration based on four recovery strategies: **a.** Betweenness, **b.** Degree, c. Closeness, and **d.** Eigenspecies, centrality. Boxplots in inset shows the fraction of species required to obtain peak MRS. MRS are computed relative to the median random recovery score (black dashed line). The best centrality based recovery strategies **(a,b)** are characterized by larger peak values.

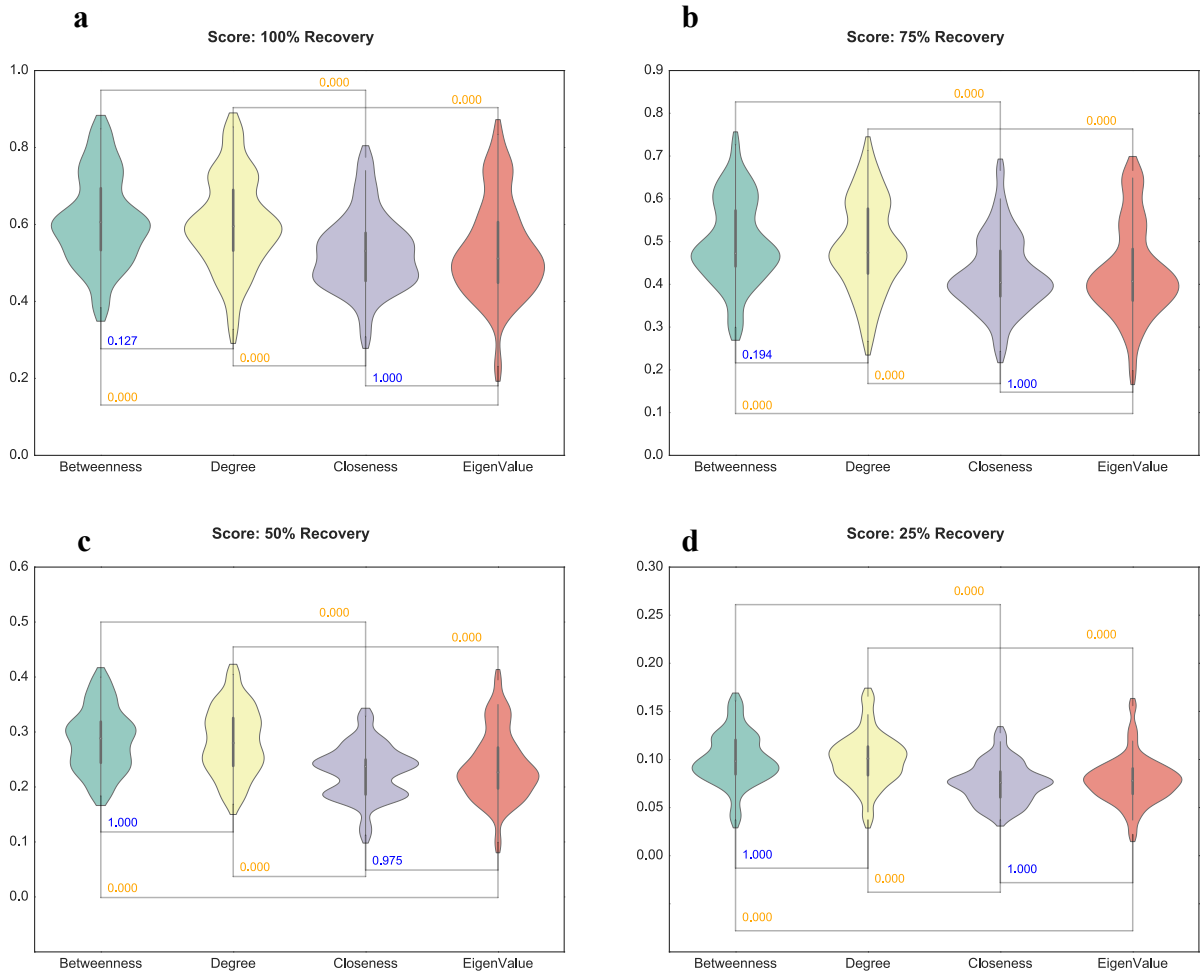

**Figure 3**| Violinplots showing the distribution in peak MRS for **a.** 100%, **b.** 75%, **c.** 50%, and **d.** 25% perturbation scenarios. Wilcoxon test with Bonferroni correction for multiple comparisons were used to determine which pairs of restoration strategies were significantly different with respect to MRS (p-value < 0.05). Significantly different restoration strategies are labeled in orange color.

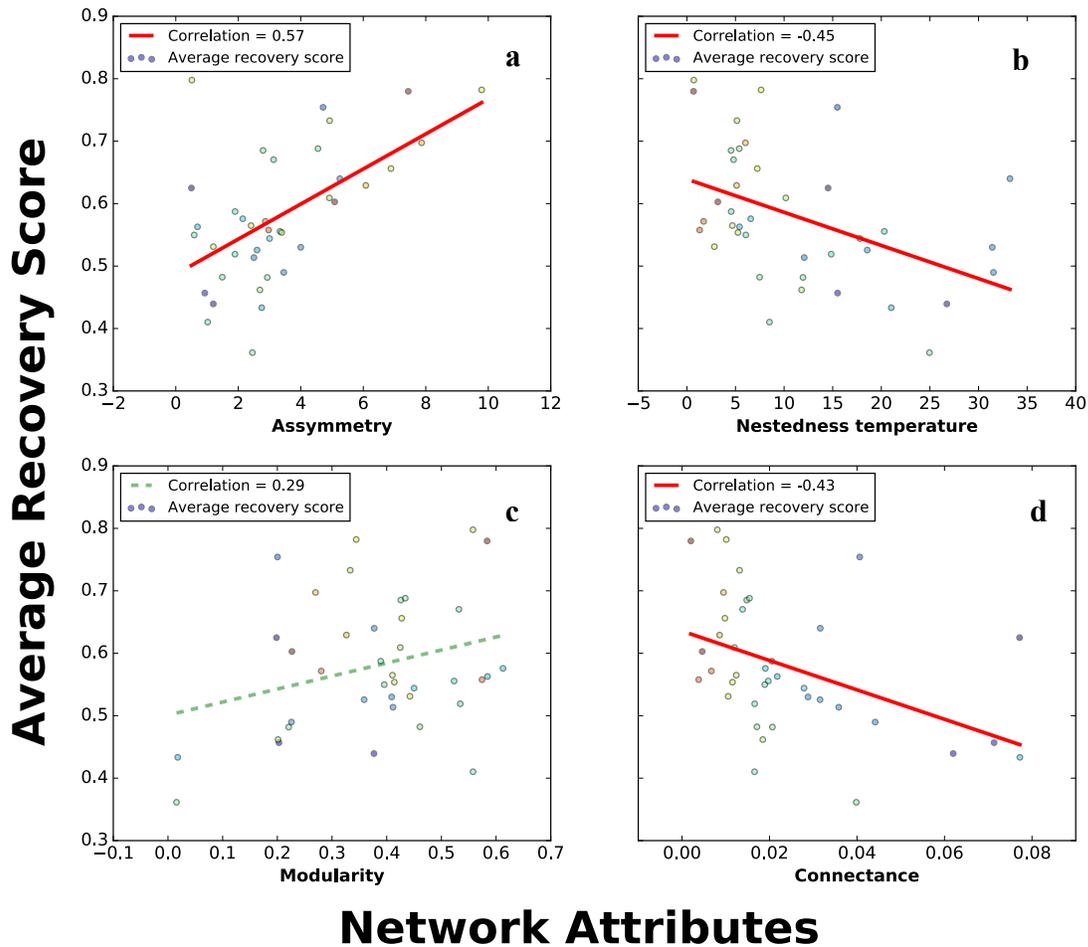

**Figure 4**| Exploratory analysis reveals novel and non-intuitive relations among restoration effectiveness metrics and ecosystem network attributes. In particular, **(A)** higher asymmetry (defined as the ratio of pollinators to plants) and **(B)** nestedness (inversely related to nestedness temperature) promote faster recovery rates. While no significant trend is observed between MRS and modularity for full restoration **(C),** connectance has an inverse effect on gains in biodiversity **(D).** Significant trends (p-values <0.05) are indicated by solid red lines.

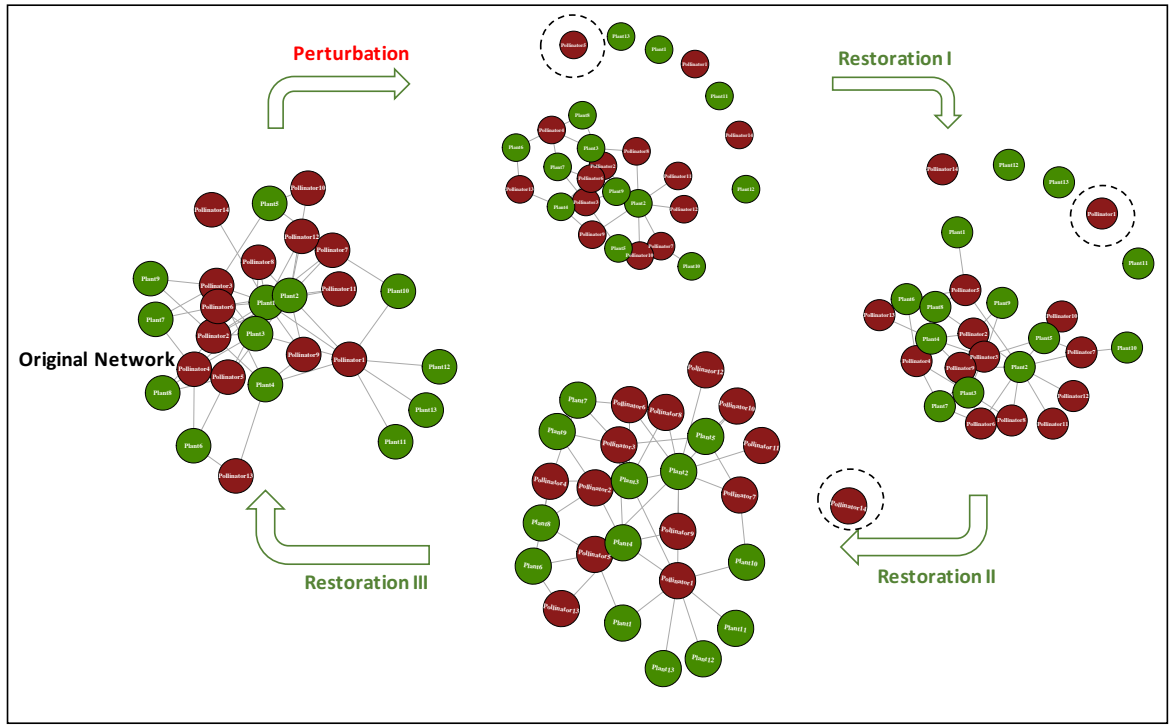

**Extended Data Figure 1|** Perturbation of ecosystem networks under external stress, specifically complete or partial species removal, and two-step restoration strategy for strategic recovery of perturbed networks. Nodes in green represent plants and those in brown represent pollinators.

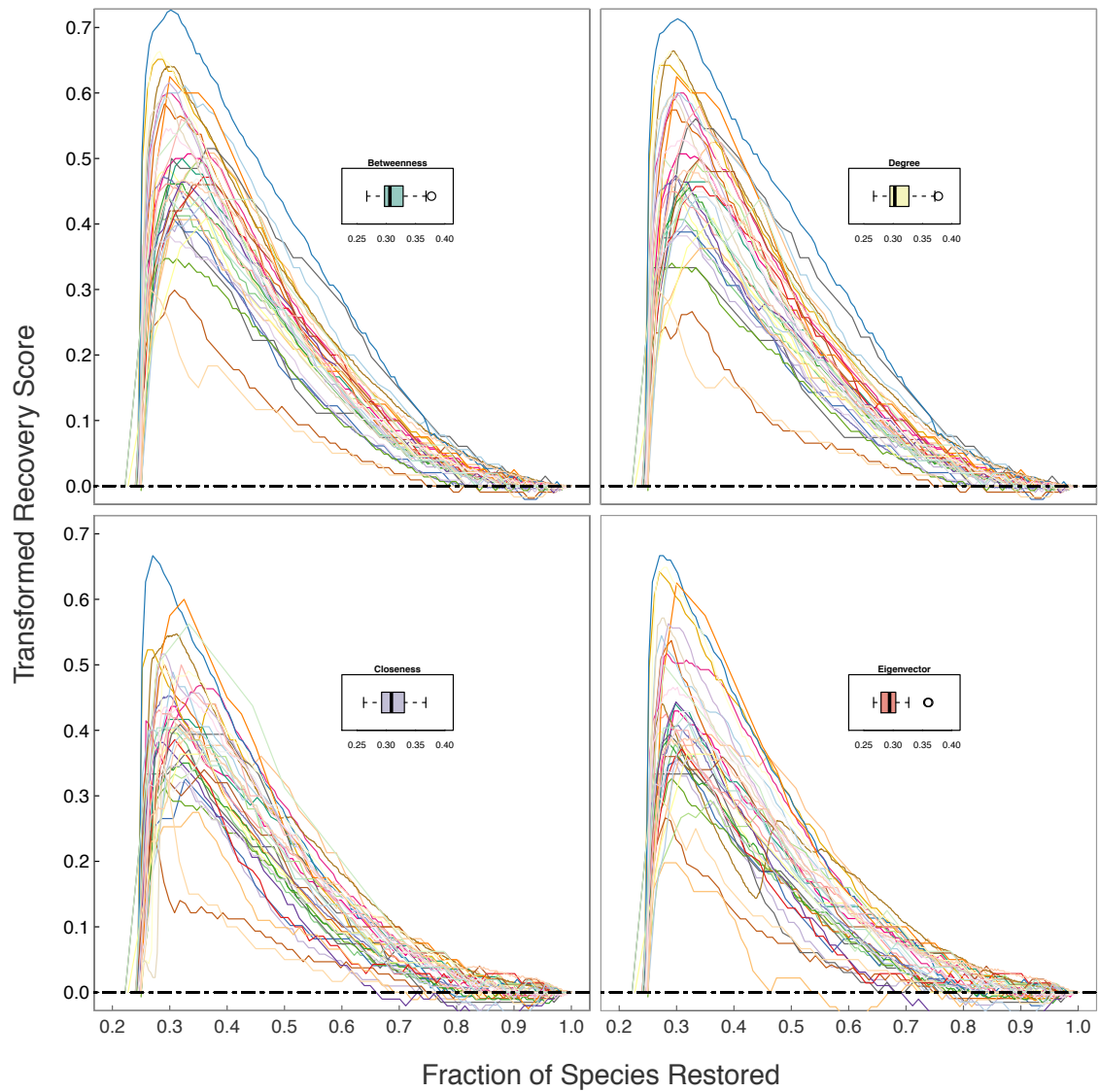

**Extended Data Figure 2|** Marginal Recovery Scores (MRS) for the four recovery strategies (Fig. 2), after partial perturbations defined as 75% species loss, based on: **a.** Betweenness, **b.** Degree, **c.** Closeness, and **d.** Eigenvector, centrality. Boxplots in inset shows the fraction of species required to obtain peak MRS. The MRS are computed relative to median random recovery scores shown by the black dashed line. Optimal recovery strategies **(a.b)** are characterized by larger peak values.

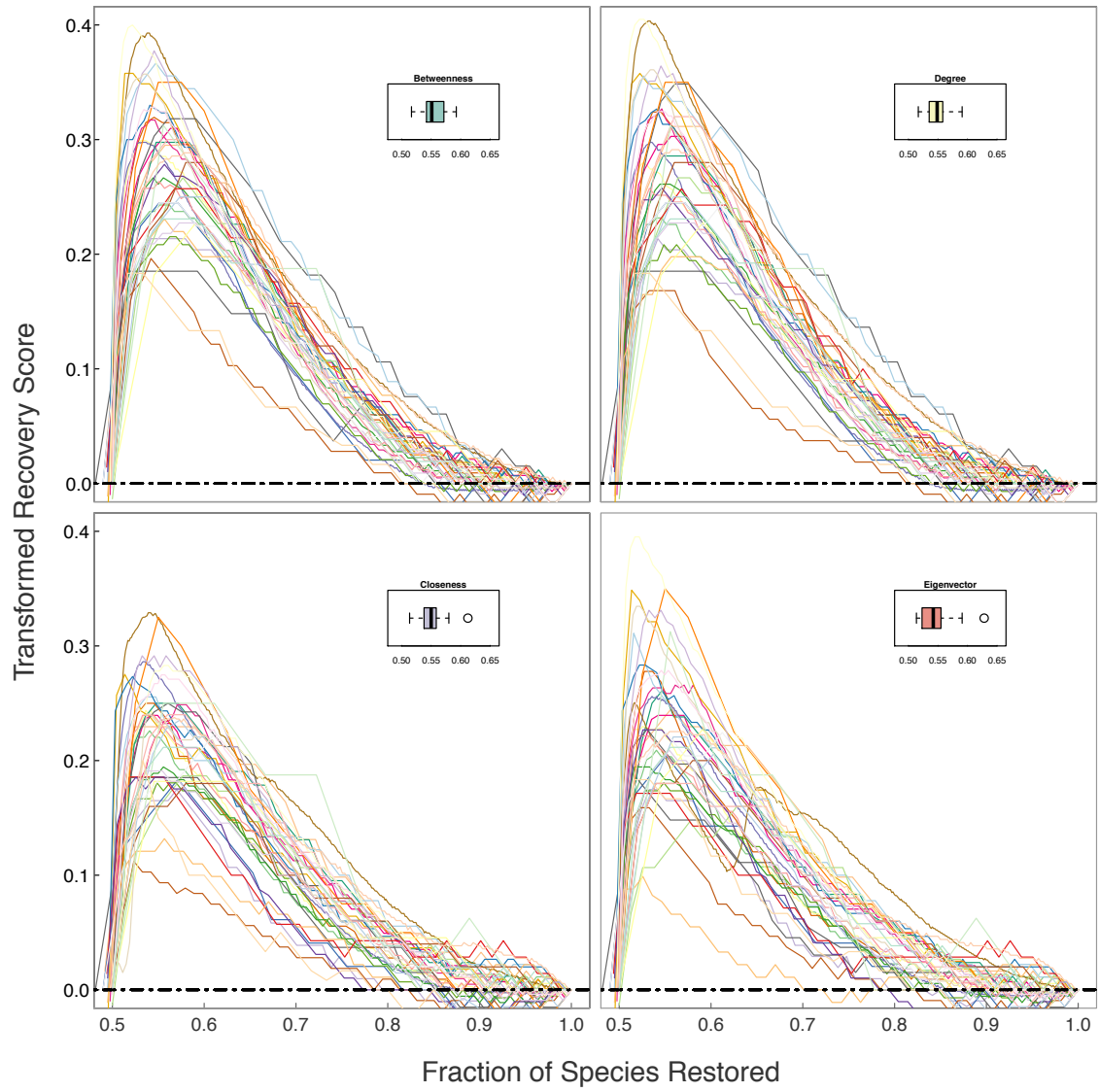

**Extended Data Figure 3|** Same as S2 but for recovery after partial perturbations defined as 50% species loss.

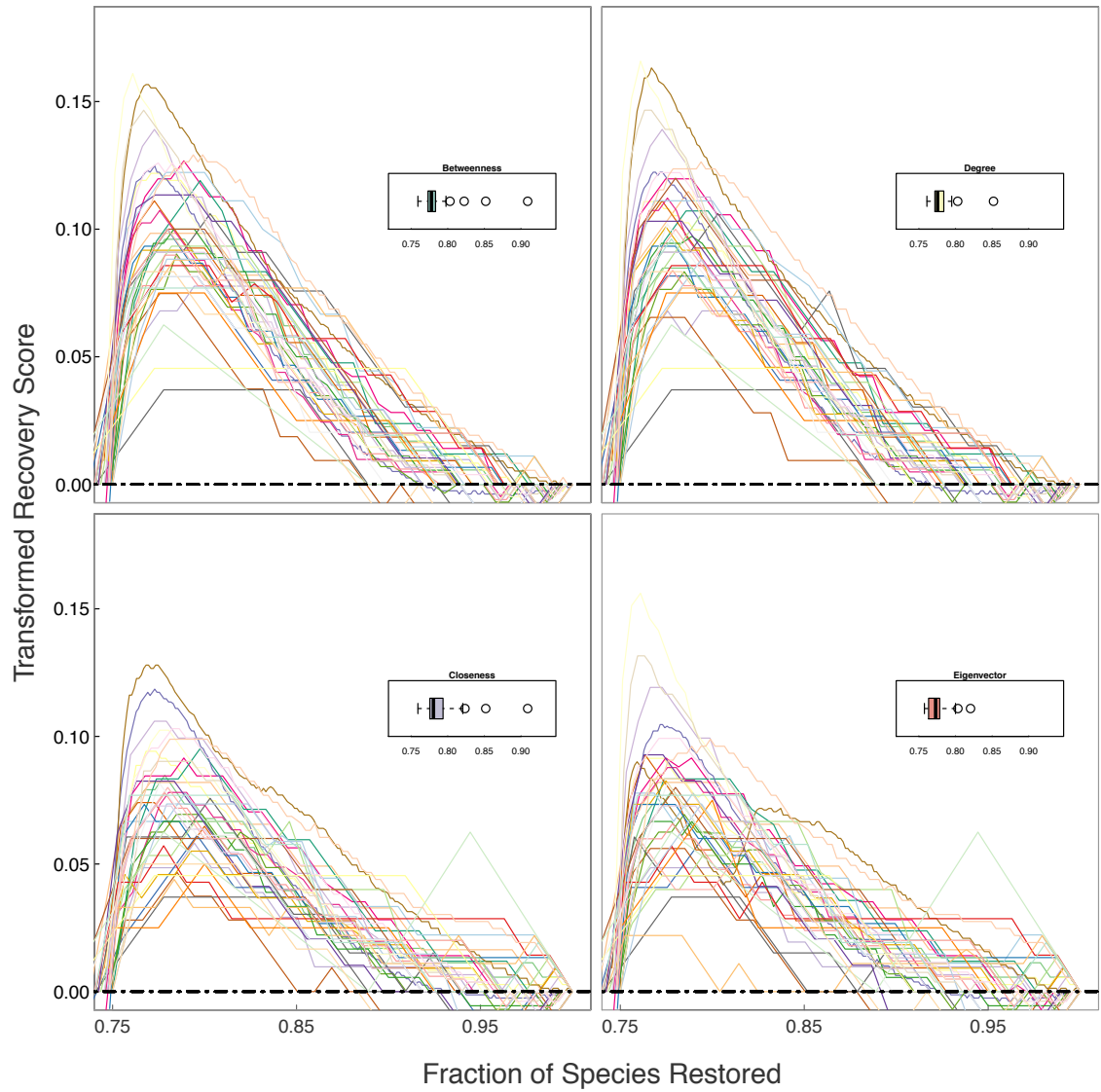

**Extended Data Figure 4|** Same as S2 but for recovery after partial perturbations defined as 25% species loss.

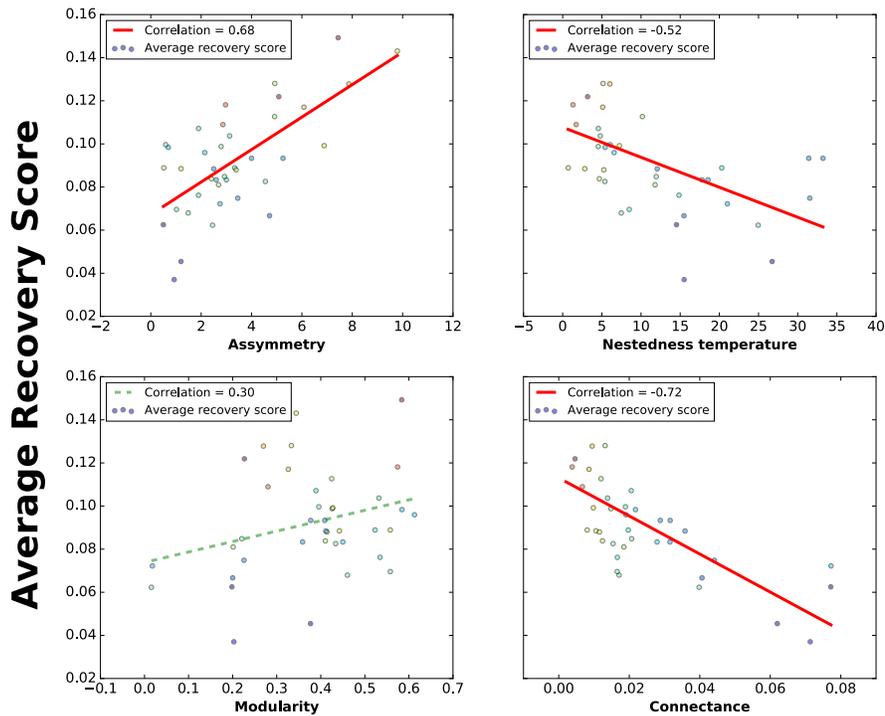

**Extended Data Figure 5|** Relationship between network attributes and marginal average recovery scores after 75% perturbation. In particular, **a.** higher asymmetry (defined as ratio of pollinators to plants) and **b.** nestedness (inversely related to nestedness temperature) promote faster recovery rates. While no significant trend is observed between MRS and modularity for partial restoration **c.,** connectance has an inverse effect on gains in biodiversity **d.** Significant trends (p-values <0.05) are indicated by solid red lines.

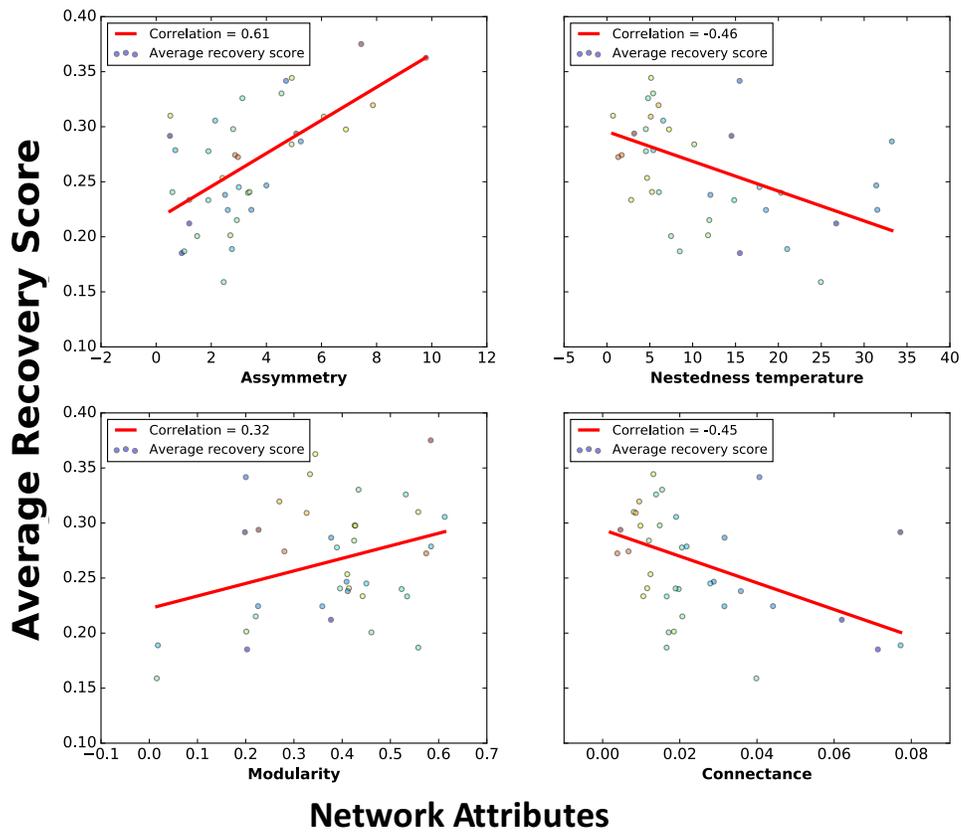

**Extended Data Figure 6|** Same as figure S5 but for recovery after 50% perturbation.

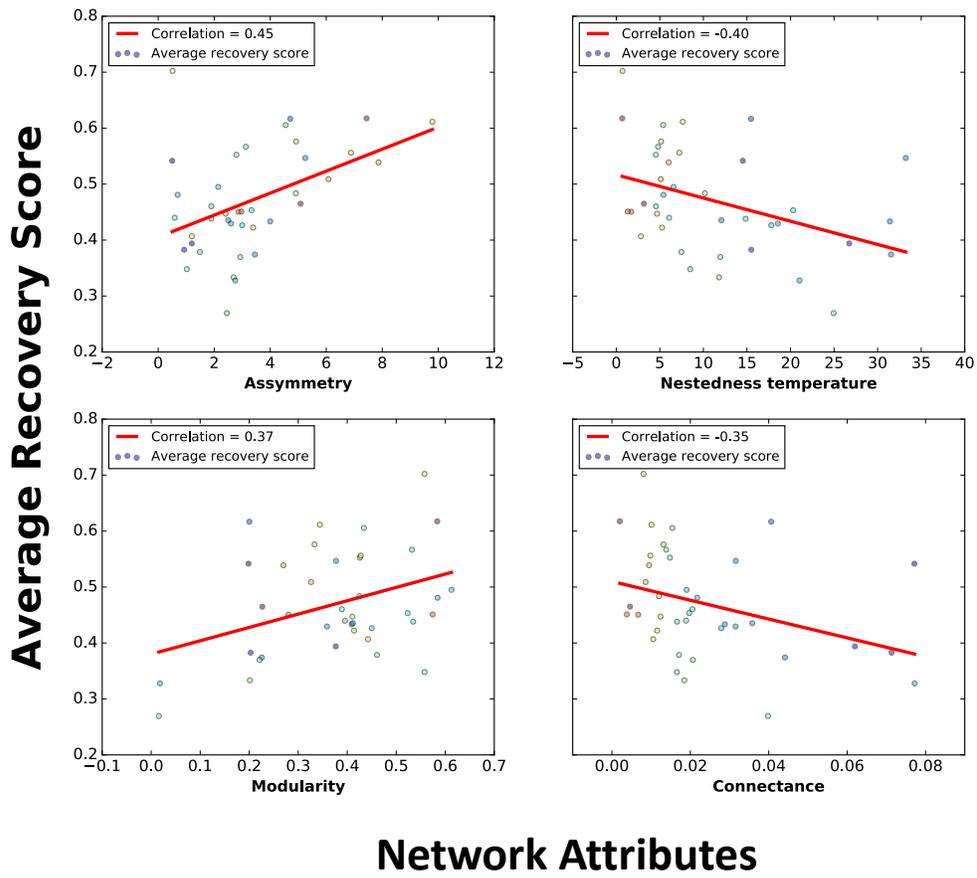

**Extended Data Figure 7|** Same as figure S5 but for recovery after 25% perturbation.

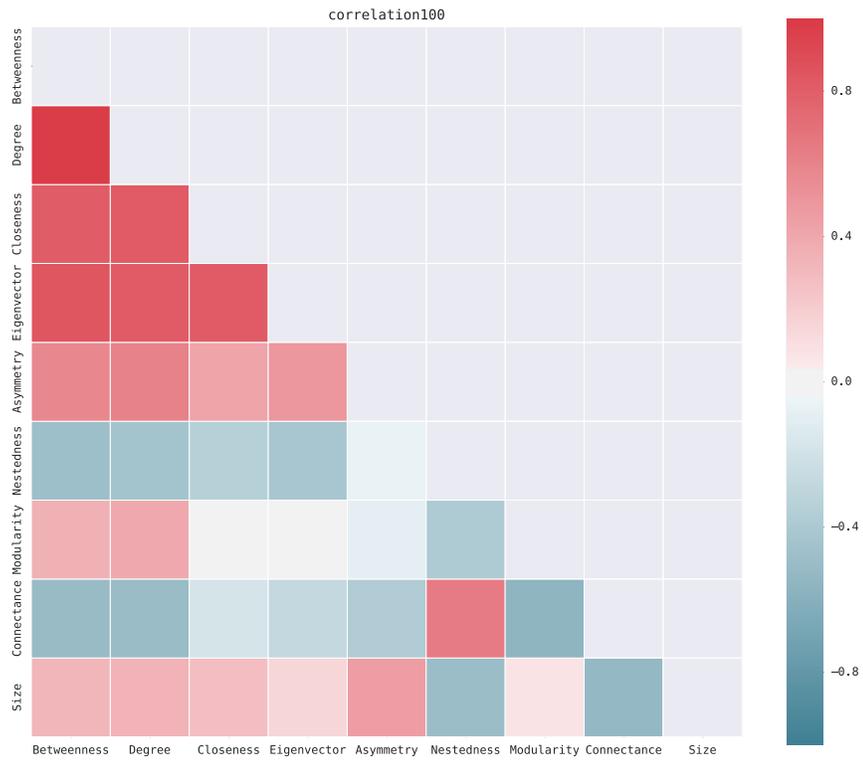

**Extended Data Figure 8**| Summary of correlations between network attributes and recovery scores for full recovery.

| Locality of Studies | Species | Interactions | Latitude | Longitude |
|---|---|---|---|---|
| Chile | 185 | 361 | -33.283334 | -70.266668 |
| Chile | 107 | 196 | -33.283334 | -70.266668 |
| Chile | 61 | 81 | -33.283334 | -70.266668 |
| Pikes Peak, Colorado, USA | 371 | 923 | 38.841784 | -105.043821 |
| Tenerife, Canary Islands | 49 | 106 | 28.216667 | -16.633333 |
| Latnjajaure, Abisko, Sweden | 142 | 242 | 68.35 | 18.5 |
| Zackenberg | 107 | 456 | 74.5 | -20.5 |
| Mauritius Island | 27 | 52 | -20.348404 | 57.552152 |
| Garajonay, Gomera, Spain | 84 | 145 | 28.127842 | -17.248908 |
| Hazen Camp, Ellesmere Island, Canada | 110 | 179 | 81.816667 | -71.3 |
| Daphn, Athens, Greece | 797 | 2933 | 38.014466 | 23.635043 |
| Doana Nat. Park, Spain | 205 | 412 | 37.016667 | -6.55 |
| Hestehaven, Denmark | 144 | 383 | 56.238737 | 9.973652 |
| Hazen Camp, Ellesmere Island, Canada | 111 | 190 | 81.816667 | -71.3 |
| Ashu, Kyoto, Japan | 768 | 1193 | 35.333333 | 135.75 |
| Laguna Diamante, Mendoza, Argentina | 66 | 83 | -34.166667 | -69.7 |
| Rio Blanco, Mendoza, Argentina | 95 | 125 | -33 | -69.283333 |
| Galapagos | 159 | 204 | -0.5 | -90.5 |
| Arthur's Pass, New Zealand | 78 | 120 | -42.95 | 171.566667 |
| Cass, New Zealand | 180 | 374 | -43.02823 | 171.78466 |
| Craigieburn, New Zealand | 167 | 346 | -43.099531 | 171.720224 |
| Guarico State, Venezuela | 81 | 109 | 8.933333 | -67.416667 |
| Canaima Nat. Park, Venezuela | 97 | 156 | 5.583333 | -61.716667 |
| Brownfield, Illinois, USA | 40 | 65 | 40.133333 | -88.166667 |
| Chiloe, Chile | 154 | 312 | -42 | -73.583333 |
| Morant Point, Jamaica | 97 | 178 | 17.916667 | -76.191667 |
| Flores | 22 | 30 | 39.466667 | -31.227222 |
| Hestehaven, Denmark | 50 | 72 | 56.248737 | 9.973652 |
| Hestehaven, Denmark | 50 | 79 | 56.238737 | 9.993652 |
| Isabela Island, Galapagos | 18 | 25 | -0.949194 | -90.977222 |
| Hestehaven, Denmark | 110 | 250 | 56.238737 | 9.953652 |
| Denmark | 60 | 278 | 56.066667 | 10.233333 |
| Isenbjerg | 205 | 425 | 56.066667 | 9.266667 |
| Denmark | 266 | 671 | 56.1 | 9.1 |
| Denmark | 262 | 590 | 56.066667 | 10.216667 |
| Tenerife, Canary Islands | 49 | 86 | 28.268611 | -16.605556 |
| Tundra, Greenladn | 54 | 92 | 66.966667 | -50.55 |
| Mt. Yufu, Japan | 393 | 589 | 33.4 | 131.5 |
| Tenerife, Canary Islands | 68 | 129 | 28.268611 | -16.605556 |

**Extended Data Table 1|** Description of the ecological networks used